\documentclass{elsart}

\usepackage{adnd}
\usepackage{longtable}

\usepackage{mathptmx}

\usepackage{amsmath}

\usepackage{amssymb,amstext}
\usepackage[pdftex,
 letterpaper=true,
 hyperindex=true,
 breaklinks=true,
 colorlinks=false,
 citecolor=blue,
 pdftitle={},
 pdfauthor={}]
{hyperref}

\usepackage{graphicx}

\begin{document}

\begin{frontmatter}

\journal{Atomic Data and Nuclear Data Tables}

\copyrightholder{Elsevier Science}

\runtitle{Barium}
\runauthor{Shore}


\title{Discovery of the Barium Isotopes}


\author{A. Shore},
\author{A. Fritsch},
\author{J.Q. Ginepro},
\author{M. Heim},
\author{A. Schuh},
\and
\author{M.~Thoennessen\corauthref{cor}},\corauth[cor]{Corresponding author.}\ead{thoennessen@nscl.msu.edu}

\address{National Superconducting Cyclotron Laboratory and \\ Department of Physics and Astronomy, Michigan State University, \\East Lansing, MI 48824, USA}

\date{July 23, 2009} 

\begin{abstract}
Thirty-eight barium isotopes have so far been observed; the discovery of these isotopes is discussed.  For each isotope a brief summary of the first refereed publication, including the production and identification method, is presented.
\end{abstract}

\end{frontmatter}





\newpage
\tableofcontents
\listofDtables

\vskip5pc

\section{Introduction}\label{s:intro}

The tenth paper in the series of the discovery of isotopes, the discovery of the barium isotopes is discussed. Previously, the discoveries of cerium \cite{Gin09}, arsenic \cite{Sho09a}, gold \cite{Sch09a}, tungsten \cite{Fri09}, krypton \cite{Hei09}, einsteinium \cite{Bur09}, iron \cite{Sch09b}, vanadium \cite{Sho09b}, and silver \cite{Sch09c} isotopes were discussed.  The purpose of this series is to document and summarize the discovery of the isotopes. Guidelines for assigning credit for discovery are (1) clear identification, either through decay-curves and relationships to other known isotopes, particle or $\gamma$-ray spectra, or unique mass and Z-identification, and (2) publication of the discovery in a refereed journal. The authors and year of the first publication, the laboratory where the isotopes were produced as well as the production and identification methods are discussed. When appropriate, references to conference proceedings, internal reports, and theses are included. When a discovery included a half-life measurement the measured value is compared to the currently adopted value taken from the NUBASE evaluation \cite{Aud03} which is based on the ENSDF database \cite{ENS08}. In cases where the reported half-life differed significantly from the adopted half-life (up to approximately a factor of two), we searched the subsequent literature for indications that the measurement was erroneous. If that was not the case we credited the authors with the discovery in spite of the inaccurate half-life.

\section{Discovery of $^{114-151}$Ba}
Thirty-eight barium isotopes  from A = $114-151$ have been discovered so far; these include seven stable, 18 proton-rich and 13 neutron-rich isotopes. According to the HFB-14 model \cite{Gor07}, $^{181}$Ba should be the last particle-stable odd-even neutron-rich nucleus, with the even-even barium isotopes reaching up to $^{188}$Ba. Along the proton dripline two more isotopes are predicted to be stable and it is estimated that five additional nuclei beyond the proton dripline could live long enough to be observed \cite{Tho04}. Thus, about 42 isotopes have yet to be discovered and almost 50\% of all possible barium isotopes have been produced and identified so far.

\begin{figure}
	\centering
	\includegraphics[width=12cm]{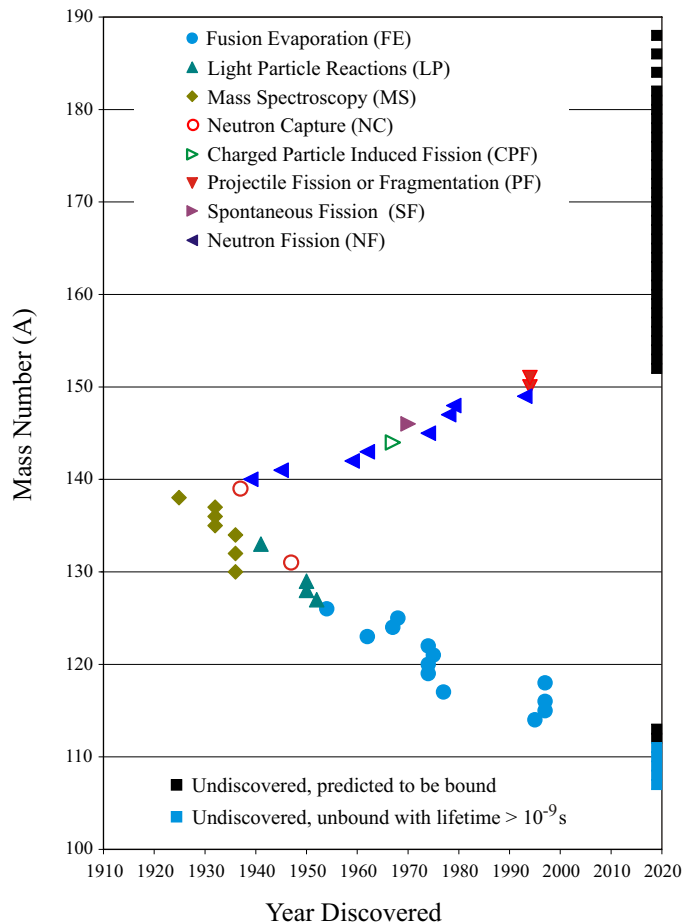}
	\caption{Barium isotopes as a function of time when they were discovered. The different production methods are indicated. The solid black squares on the right hand side of the plot are isotopes predicted to be bound by the HFB-14 model. On the proton-rich side the light blue squares correspond to unbound isotopes predicted to have lifetimes larger than $\sim 10^{-9}$~s.}
	\label{f:year}
\end{figure}

Figure \ref{f:year} summarizes the year of first discovery for all barium isotopes identified by the method of discovery. The range of isotopes predicted to exist is indicated on the right side of the figure. The radioactive barium isotopes were produced using heavy-ion fusion evaporation (FE), light-particle reactions (LP), neutron-capture reactions (NC), spontaneous fission (SF), neutron induced fission (NF), charged-particle induced fission (CPF), and projectile fragmentation or fission (PF). The stable isotopes were identified using mass spectroscopy (MS). Heavy ions are all nuclei with an atomic mass larger than A = 4 \cite{Gru77}. Light particles also include neutrons produced by accelerators. In the following paragraphs, the discovery of each barium isotope is discussed in detail.

\subsection*{$^{114}$Ba}\vspace{-.85cm}
In 1995 Guglielmetti \textit{et al.} announced in \textit{Identification of the new isotope $^{114}$Ba and search for its $\alpha$ and cluster radioactivity} the discovery of $^{114}$Ba \cite{Gug95a}. $^{114}$Ba was produced at the Gesellschaft f\"ur Schwerionenforschung Unilac via the fusion evaporation reaction $^{58}$Ni($^{58}$Ni,2n)$^{114}$Ba at 4.2~MeV/u and identified using an on-line mass separator. ``With $\Delta$E-E telescopes we measured the total ($\beta$-decay) half-life to be T$_\beta$ = 0.43$^+{0.30}_{-0.15}$~s and the partial $\alpha$-decay half-life to be T$_\alpha \ge$ 1.2 $\times$ 10$^2$~s (1 MeV $le$ E$_\alpha \le$ 4 MeV) for $^{114}$Ba.'' This half-life agrees with the currently accepted value of 0.53(23)~s.

\subsection*{$^{115,116}$Ba}\vspace{-.85cm}
In the paper \textit{Decay studies of the neutron-deficient isotopes $^{114-118}$Ba} Janas \textit{et al.} reported the first observation of $^{115}$Ba and $^{116}$Ba in 1997 \cite{Jan97}. A 4.9 MeV/u $^{58}$Ni beam was accelerated by the linear accelerator UNILAC at GSI and bombarded enriched $^{58}$Ni and $^{60}$Ni targets. $^{115}$Ba and $^{116}$Ba were identified by measuring the energy and time of $\beta$-delayed protons following on-line mass separation. ``The least-square fit yielded $T_{1/2} = 0.45 \pm 0.05$ s for the decay half-life of $^{115}$Ba and a lower limit of 15\% for $b_{\beta p}$.'' In addition, $\beta$-delayed X-rays were detected for $^{116}$Ba. ``The time characteristics of the Cs KX-rays intensity, analyzed under the assumption of a single decay component, yielded $T_{1/2} = 1.3 \pm 0.2$ s for the half-life of $^{116}$Ba.'' These observed half-lives are currently the only measured values for $^{115}$Ba and $^{116}$Ba. The same group had previously mentioned the observation of these isotopes in a conference proceeding \cite{Gug95b}.

\subsection*{$^{117}$Ba}\vspace{-.85cm}
Bogdanov \textit{et al.} reported the discovery of $^{117}$Ba in 1977 in their article \textit{New Neutron-Deficient Isotopes of Barium and Rare-Earth Elements} \cite{Bog77}. An enriched $^{92}$Mo target was bombarded with an 180-190~MeV sulfur beam produced by the JINR Laboratory of Nuclear Reactions U-300 Heavy Ion Cyclotron and $^{117}$Ba was produced in the fusion evaporation reaction $^{92}$Mo($^{32}$S,2p5n). The isotope was separated with the BEMS-2 on-line ion source and identified by its delayed proton emission. ``As the proton emission of the A = 117 isobar, it is unambiguously related to $^{117}$Ba, since the reaction leading to the formation of $^{117}$Cs was energetically impossible in our experiments.'' The measured half-life of 1.9(2)~s is consistent with the accepted value of 1.75(7)~s.

\subsection*{$^{118}$Ba}\vspace{-.85cm}
In the 1997 paper \textit{Decay studies of the neutron-deficient isotopes $^{114-118}$Ba} Janas \textit{et al.} reported the first observation of $^{118}$Ba \cite{Jan97}. A 4.9 MeV/u $^{58}$Ni beam was accelerated by the linear accelerator UNILAC at GSI and bombarded enriched $^{58}$Ni and $^{60}$Ni targets on a $^{63}$Cu backing. $^{118}$Ba was produced in the fusion-evaporation reaction $^{63}$Cu($^{58}$Ni,1p2n) and identified by measuring $\beta$-delayed X-rays and $\gamma$-rays following on-line mass separation. ``From the time characteristics of the cesium KX-rays intensity the $^{118}$Ba half-life of $T_{1/2} = 5.2 \pm 0.2$~s was determined under the assumption of a single decay component.'' This observed half-life is currently the only measured value for $^{118}$Ba. The same group had previously mentioned the observation of this isotope in a conference proceeding \cite{Gug95b}.

\subsection*{$^{119}$Ba}\vspace{-.85cm}
In 1974 Bogdanov \textit{et al.} observed $^{119}$Ba, which they reported in the article \textit{Delayed-proton emitter $^{119}$Ba} \cite{Bog74}. An enriched $^{106}$Cd target was bombarded with an 85~MeV oxygen beam produced by JINR Laboratory of Nuclear Reactions U-300 Heavy Ion Cyclotron. $^{119}$Ba was produced in the fusion-evaporation reaction $^{106}$Cd($^{16}$O,3n). The delayed proton spectrum was measured with a telescope of a thin proportional-counter and a surface-barrier detector. ``The observed activity was due mainly to two radiators with half-lives T$_{1/2} = 5.0 \pm 0.6$ sec, with an excitation function peaking at E$_{^{16}O}$ = 85 MeV, and T$_{1/2} = 15.0 \pm 1.0 $ sec with its maximum yield at a higher energy... Thus, the most probable identification of the 5-second emitter is $^{119}$Ba.'' The extracted half-life agrees with the currently accepted value of 5.4(3)~s.

\subsection*{$^{120}$Ba}\vspace{-.85cm}
In 1974 Conrad \textit{et al.} reported the observation of $^{120}$Ba in their article \textit{Quasi-Rotational Bands in Neutron Deficient Doubly Even Ba Isotopes} \cite{Con74}. $^{120}$Ba was produced in the fusion evaporation reaction $^{106}$Cd($^{16}$O,2n) by bombarding cadmium with a 66~MeV oxygen beam provided by the MP Tandem of the Max-Planck-Institut f\"{u}r Kernphysik in Heidelberg, Germany. The isotope was identified by charged-particle-, neutron-, and gamma-gamma coincidence measurements: ``To identify $^{120}$Ba, which has 18 neutrons less than the most abundant barium isotope, neutron-gamma coincidences had to be applied in addition to charged particle and gamma-gamma coincidence measurements. The upper limit for the lifetime of the ground state of $^{120}$Ba is 90~sec.'' The first three $\gamma$-transitions in $^{120}$Ba were measured. The upper limit of 90~seconds is consistent with the accepted half-life value of 24(2)~s.

\subsection*{$^{121}$Ba}\vspace{-.85cm}
In the 1975 paper \textit{New Delayed-Proton Emitters $^{119}$Ba, $^{121}$Ba and $^{116}$Cs} Bogdanov \textit{et al.} reported the discovery of $^{121}$Ba \cite{Bog75}. The U-300 cyclotron of the Nuclear Reactions Laboratory at Dubna accelerated a $^{32}$S beam to a maximum energy of 190~MeV. ``The isotope $^{121}$Ba was observed in bombardment of niobium with sulfur ions in the reaction $^{93}$Nb($^{32}$S,p3n)$^{121}$Ba and with substantially greater yield in the reaction $^{92}$Mo($^{32}$S,2pn). The half-life is 29.7$\pm$1.5~sec.'' $^{121}$Ba was identified with the BEMS-2 mass separator. The observed half-life is currently the only available measurement. A month later the authors submitted the results to a different journal where they were published first \cite{Kar74}.

\subsection*{$^{122}$Ba}\vspace{-.85cm}
In 1974 Conrad \textit{et al.} reported the observation of $^{122}$Ba in their article \textit{Quasi-Rotational Bands in Neutron Deficient Doubly Even Ba Isotopes} \cite{Con74}. $^{122}$Ba was produced in the fusion evaporation reaction $^{108}$Cd($^{16}$O,2n) by bombarding cadmium with a 66~MeV oxygen beam provided by the MP Tandem of the Max-Planck-Institut f\"{u}r Kernphysik in Heidelberg, Germany. The isotope was identified by gamma-gamma coincidence measurements: ``Up to now a level scheme of $^{122}$Ba has not been published. For the lifetime of the ground state a value between 2.5 and 5 sec has been suggested \cite{DAu67}. From our data the partial level scheme shown in fig. 4 was obtained.'' The first six $\gamma$-transitions of the ground state band in $^{122}$Ba were measured.

\subsection*{$^{123}$Ba}\vspace{-.85cm}
Preiss and Strudler reported discovery of $^{123}$Ba in 1962 in their article \textit{New Neutron Deficient Barium Isotopes} \cite{Pre62}. $^{123}$Ba was produced via the fusion-evaporation reactions $^{113}$In($^{16}$O,p5n), $^{115}$In($^{16}$O,p7n), $^{113}$In($^{14}$N,4n), $^{115}$In($^{14}$N,6n), natural Sn($^{16}$O,$\alpha$xn)$^{123}$Ba and natural Sn($^{12}$C,xn)$^{123}$Ba; the beams were produced by the Yale University Heavy Ion Accelerator and had a maximum energy of 10.5~MeV/nucleon. $^{123}$Ba was identified measuring characteristic X-ray spectra following chemical separation. ``Mass assignments for the new Ba activities were based on the parent daughter genetics using Cs half-lives and $\gamma$-ray energies previously reported and/or found in the present study. The proposed half-lives and mass assignments are: $^{123}$Ba, 2$\pm$0.5 min; $^{125}$Ba, 6.5$\pm$0.5~min; and $^{127}$Ba, 10.0$\pm$0.5~min.'' The observed half-life for $^{123}$Ba is consistent with the currently accepted value of 2.7(4)~m.

\subsection*{$^{124}$Ba}\vspace{-.85cm}
$^{124}$Ba was first observed in 1967 by Clarkson \textit{et al.} as reported in \textit{Collective Excitations in Neutron-Deficient Barium, Xenon, and Cerium Isotopes} \cite{Cla67}. $^{124}$Ba was produced by the reactions $^{115}$In($^{14}$N,5n) at 84~MeV and $^{116}$Sn($^{12}$C,4n) at 80~MeV where the ions were accelerated with the Berkeley heavy-ion linear accelerator~(HILAC). Gamma-ray spectra were measured with a lithium-drifted germanium counter. ``Since $^{124}$Ba was produced by two reactions with different targets and projectiles, which both give the same transitions, this mass assignment is likewise considered to be quite certain.'' The first two transitions of the ground-state band were correctly identified.

\subsection*{$^{125}$Ba}\vspace{-.85cm}
$^{125}$Ba was identified correctly for the first time in 1968 by D'Auria \textit{et al.} in the article \textit{Deformation in the Light Ba Isotopes: Isomeric States of Ba$^{125}$ and Ba$^{127}$} \cite{DAu68}. $^{125}$Ba was produced in the fusion-evaporation reaction $^{115}$In($^{14}$N,4n) at Yale University. The isotope was chemically separated and $\beta$- $\gamma$- and X-ray spectra were measured. ``Previously unobserved and unassigned $\gamma$ rays resulting from the decay of Ba$^{125}$ were detected at 56$\pm$3, 76$\pm$2, 84$\pm$2, and 141$\pm$2 keV, decaying with a composite half-life of 3.0$\pm$0.5 min.'' This half-life agrees with the currently accepted value of 3.5(4)~m.
D'Auria \textit{et al.} interpreted this state as the high-spin ground state in addition to the presence of an 8(1)~m isomeric excited state. The existence of the isomeric state has not been confirmed. Preiss and Strudler had previously reported a half-life of 6.5(5)~m for $^{125}$Ba \cite{Pre62}. However, because this value is almost a factor of two larger than the accepted value and closer to the claimed isomeric state, it is likely that the state had been misidentified.

\subsection*{$^{126}$Ba}\vspace{-.85cm}
In the 1954 article \textit{New Chain Barium-126--Cesium-126} Kalkstein \textit{et al.} announced the discovery of $^{126}$Ba \cite{Kal54}. Indium oxide was bombarded with a nitrogen beam produced by the Berkeley Crocker 60~inch cyclotron with a maximum energy of 140~MeV. $^{126}$Ba was produced in the fusion-evaporation reaction $^{115}$In($^{14}$N,3n) and identified using a scintillation spectrometer, a scintillation coincidence spectrometer, and a time-of-flight mass spectrograph. Element assignments and genetic relations have been verified chemically, and the mass number assigned with the isotope separator. ``$^{126}$Ba decays principally by orbital electron capture with a half-life of 96.5$\pm$2.0~minutes.'' This half-life is consistent with the currently accepted value of 100(2)~m.

\subsection*{$^{127}$Ba}\vspace{-.85cm}
In 1952 Linder and Osborne reported the discovery of $^{127}$Ba in \textit{The Nuclides Ba$^{127}$, Ba$^{128}$ and Cs$^{128}$} \cite{Lin52}. A cesium nitrate target was bombarded with 190~MeV deuterons at Livermore. $^{127}$Ba was chemically separated and its activity measured with an end-window argon-alcohol-filled counter. ``A barium isotope of 12-minute half-life was found whose radiations were not directly characterized. However, positron emission is probable since electromagnetic radiation seemed to comprise no more than five percent of the total activity detectable on an end-window counter. By four rapid chemical separations made at ten-minute intervals a cesium activity was obtained from the barium whose half-life and radiation characteristics agree with those reported for Cs$^{127}$. Furthermore, the yield of this nuclide diminished roughly by a factor of two in each of the four successive separations. The 12-minute barium activity is thus Ba$^{127}$.'' The observed half-life agrees with the currently adopted value of 12.7(4)~m.

\subsection*{$^{128}$Ba}\vspace{-.85cm}
Fink and Templeton identified $^{128}$Ba in 1950 as described in \textit{Radioactive Isotopes of Barium} \cite{Fin50}. Cesium chloride was bombarded with 85~MeV protons in the 184~inch Berkeley cyclotron. The induced barium activities were chemically separated and the half-life was measured with a Geiger counter: ``The 2.4-day period is not $^{127}$Ba, otherwise it would produce 5.5-hour $^{127}$Cs as a daughter. The most probable assignment is $^{128}$Ba from the p,6n reaction, but this assignment lacks direct proof.'' The measured half-life of 2.4~d agrees with the presently accepted value of 2.43(5)~d. Thomas and Wiig had measured a 2.4(1)~d half-life in barium but they were only able to determine that the mass number was smaller than 129 \cite{Tho50}.

\subsection*{$^{129}$Ba}\vspace{-.85cm}
In 1950 the discovery of $^{129}$Ba was reported first by Thomas and Wiig in \textit{On Neutron Deficient Isotopes of Barium} \cite{Tho50}.
250-MeV protons accelerated by the Rochester 130-inch cyclotron were used to bombard spectroscopically pure cesium chloride. The half-life of the chemically separated barium fraction was measured by deflecting positrons into a counter tube with a permanent magnet. ``Parent-daughter separations performed more than 24 hours after the bombardment failed to show any $^{129}$Cs activity whereas earlier milkings did show the activity. This led to the conclusion that the 1.8-hour barium is $^{129}$Ba.'' The measured half-life of 1.8(2)~h is close to the currently accepted value of 2.23(11)~h. Less than a month later Fink and Templeton submitted their half-life measurement of 2.0(1)~h for $^{129}$Ba \cite{Fin50} which was published in the same issue of Physical Review immediately following the article by Thomas and Wiig.

\subsection*{$^{130}$Ba}\vspace{-.85cm}
In the 1936 paper \textit{The Isotopic Constitution of Barium and Cerium} Dempster reported the first observation of $^{130}$Ba \cite{Dem36}. The mass spectra were measured at the Ryerson Physical Laboratory at the University of Chicago: ``I have photographed several mass spectra of the barium ions formed in a high frequency spark between two barium electrodes, which show two still fainter isotopes at 130 and 132.''

\subsection*{$^{131}$Ba}\vspace{-.85cm}
$^{131}$Ba was first observed by Katcoff in 1945 and reported in Plutonium Project Records \cite{Kat45}. The results were subsequently published in a refereed journal in the article \textit{New Barium and Cesium Isotopes: 12.0d Ba$^{131}$, 10.2 Cs$^{131}$, and Long-Lived Ba$^{133}$} \cite{Kat47}. $^{131}$Ba was produced by neutron irradiation of BaCO$_{3}$ in the Clinton Pile of Argonne National Laboratory and separated via fractional precipitation experiments. ``The Ba$^{131}$ isotope decays predominantly by orbital electron capture with a half-life of 12.0 days, emitting gamma-radiations of about 0.26 Mev, 0.5 Mev, and roughly 1.2 Mev.'' This half-life agrees with the presently adopted value of 11.50(6)~d. Yu \textit{et al.} had submitted their observation of a 11.7(3)~d half-life \cite{Yu47} for $^{131}$Ba six months before Katcoff. Although they do not reference the Plutonium Project Records by Katcoff they utilized material made available by the Manhattan Project \cite{Man46}. Thus we still credit Katcoff with the first observation of $^{131}$Ba.

\subsection*{$^{132}$Ba}\vspace{-.85cm}
In the 1936 paper \textit{The Isotopic Constitution of Barium and Cerium} Dempster reported the first observation of $^{132}$Ba \cite{Dem36}. The mass spectra were measured at the Ryerson Physical Laboratory at the University of Chicago: ``I have photographed several mass spectra of the barium ions formed in a high frequency spark between two barium electrodes, which show two still fainter isotopes at 130 and 132.''

\subsection*{$^{133}$Ba}\vspace{-.85cm}
Cork and Smith reported the first observation of $^{133}$Ba in their 1941 article \textit{Radioactive Isotopes of Barium from Cesium} \cite{Cor41}. $^{133}$Ba was produced by bombarding cesium with 9.5 MeV deuterons at the University of Michigan. Following chemical separation the isotope was identified with a magnetic beta-spectrometer and by absorption measurements. The observed 40.0(5)~h half-life was identified as an excited state and could be ascribed to either $^{133}$Ba or $^{134}$Ba. ``However, it is known that the Rochester group have found this same activity by bombarding cesium with protons. In their bombardment it is possible to produce barium by the (P,N) and (P,$\gamma$) reactions. Although the latter process it known to occur, the former is much more probable and favors the assignment to Ba$^{133}$.'' The observed half-life of this isomeric state agrees with the present value of 38.9(1)~h.

\subsection*{$^{134}$Ba}\vspace{-.85cm}
In 1936 Blewett and Sampson reported the discovery of $^{134}$Ba in \textit{Isotopic Constitution of Strontium, Barium, and Indium} \cite{Ble36}. The mass spectrographic study of barium was performed at Princeton University by heating barium oxide from a tungsten filament. ``The curves for barium showed a peak at mass 134 making up 1.8 percent of the total emission...[which] lead us to believe that this is due to a new isotope of barium.''

\subsection*{$^{135-137}$Ba}\vspace{-.85cm}
$^{135}$Ba, $^{136}$Ba, and $^{137}$Ba were discovered by Aston as reported in his 1932 article \textit{The Isotopic Constitution and Atomic Weights of Caesium, Strontium, Lithium, Rubidium, Barium, Scandium and Thallium} \cite{Ast32}. The mass spectra were measured at the Cavendish Laboratory in Cambridge, UK: ``The production of sufficiently intense rays from barium salts is a matter of great difficulty, but after many attempts an anode containing the chloride mixed with a little iodide yielded mass-spectra showing beyond any doubt the presence of three new isotopes, 135, 136, 137.''

\subsection*{$^{138}$Ba}\vspace{-.85cm}
In 1925 Aston reported the first observation of $^{138}$Ba in \textit{The Mass Spectra of Chemical Elements, Part VI. Accelerated Anode Rays Continued} \cite{Ast25}. The mass spectra were measured at the Cavendish Laboratory in Cambridge, UK: ``In these experiments the anode consisted of a mixture of barium chloride and lithium bromide. Schumannized plates were used and the line $^{138}$Ba was obtained of very great intensity.''

\subsection*{$^{139}$Ba}\vspace{-.85cm}
Pool \textit{et al.} published the first identification of $^{139}$Ba in \textit{A Survey of Radioactivity Produced by High Energy Neutron Bombardment} in 1937 \cite{Poo37a}. Neutrons with energies up to 20 MeV, produced by bombarding lithium with 6.3 MeV deuterons at the University of Michigan, were used to irradiate many stable elements. In the summary table the observed half-life of 85~m was assigned to $^{139}$Ba. This assignment is supported by a previously published contribution to a conference: ``Barium becomes strongly radioactive with a half-life period of 85.6 min. The $\beta$-particles have the negative sign. Chemical analysis shows that the activity is most probably due to Ba$^{139}$.'' \cite{Poo37b}. This half-life agrees with the currently accepted value of 83.1(3)~m. Amaldi \textit{et al.} had reported a 80~m period in barium in 1935; however, no mass assignment was made \cite{Ama35}.

\subsection*{$^{140}$Ba}\vspace{-.85cm}
In the 1939 paper \textit{Nachweis der Entstehung aktiver Bariumisotope aus Uran und Thorium durch Neutronenbestrahlung; Nachweis weiterer aktiver Bruchst\"ucke bei der Uranspaltung} Hahn and Strassmann identified $^{140}$Ba for the first time at Berlin-Dahlem in Germany \cite{Hah39}. $^{140}$Ba was produced by irradiating Uranium with neutrons from a Ra-Be-neutron source. Decay curves were measured following chemical separation. A previously reported 300~h activity originally labeled as ``Ra IV'' \cite{Hah38,Hah39a} now identified as the fission product ``Ba IV'' was again observed. Based on the measured half-life of the daughter activity it was tentatively assigned to $^{140}$Ba: ``Was die anderen Barium isotope aus dem Uran anbelangt, so l\"a\ss t sich f\"ur das Ba IV vielleicht die Hypothese machen, da\ss\ es die Muttersubstanz des in der Literatur beschriebenen Radiolanthans von 31-46 Stunden Halbwertszeit mit dem vermutlichen Atomgewicht 140 ist.'' (Concerning the other from uranium produced barium isotopes, it is hypothesized that Ba IV may be the parent of the radioactive lanthanum which was reported with a half-life of 31-46 hours with the probable atomic weight of 140.) Hahn and Strassmann did not officially assign the 300~h activity to $^{140}$Ba in subsequent papers \cite{Hah39b,Hah39c,Hah40a,Hah40b}, although they confirmed the relationship of the activity to $^{140}$La \cite{Hah40b}. Subsequently this assignment was specifically made by other authors \cite{Bor43,Wei43} and it was generally adopted in the 1944 Table of Isotopes \cite{Sea44}. The final proof was given by mass-spectroscopic measurements in 1947 \cite{Hay48}.

\subsection*{$^{141}$Ba}\vspace{-.85cm}
Katcoff established the identification of $^{141}$Ba in 1945 in the Plutonium Project Record \textit{Radiations from 3.7h La$^{141}$} \cite{Kat45b}.
Uranyl nitrate was irradiated with slow neutrons produced with the Chicago cyclotron and the A=141 mass chain of the $^{141}$Ba-$^{141}$La-$^{141}$Ce relationship was established: ``About 75 min was then allowed for 3.7h La$^{141}$ to grow into the solution from its 18m Ba$^{141}$ parent...The $\beta$-decay curve shows a long-lived component (probably the 28d Ce$^{141}$ daughter of 3.7h La$^{141}$) and small amounts of 30h and 1.5h components; but the 3.7h component greatly predominates.'' The currently accepted value for the half-life of $^{141}$Ba is 18.27(7)~m. Hahn and Strassmann had originally reported this half-life in barium for the first time \cite{Hah42} modifying a previous observation of a single 14~m component \cite{Hah39} into two components of 18~m and 6~m. Hahn and Strassmann also observed the relationship of the 18~m half-life with a 3.5~h component in lanthanum. However, no specific mass assignment was made. In another paper in the Plutonium Project Record Goldstein mentioned the established relationship of the mass chain \cite{Gol44} referring to the work by Ballou and Burgus which was not included in the published record \cite{Bal43}.

\subsection*{$^{142}$Ba}\vspace{-.85cm}
The first accurate identification of $^{142}$Ba was published by Schuman \textit{et al.} in 1959 in the article \textit{Decay of Short-Lived Barium and Lanthanum Fission Products} \cite{Sch59}. Enriched $^{235}$U was irradiated in an MTR pneumatic rabbit facility of the Atomic Energy Division of the Phillips Petroleum Company at Idaho Falls, Idaho. Fission products were chemically separated and the $\beta$- and $\gamma$- decays of the fragments were measured. ``The gamma-ray spectra of the barium samples showed photopeaks decaying with two half-lives, 11 min for Ba$^{142}$ and 18 min for Ba$^{141}$ and in addition the lanthanum daughter photopeaks growing in.'' The reported half-life of 11(1)~m agrees with the presently accepted value of 10.6(2)~m. The 6~m half-life reported by Hahn and Strassmann in 1942 \cite{Hah42} was only tentatively assigned to $^{142}$Ba as late as the 1958 edition of the Table of Isotopes \cite{Str58} (Classification D: Element certain and mass number not well established). Mal\'y \textit{et al.} reported a value of 5.9~m for $^{142}$Ba in 1958 \cite{Mal58}. Due to the large discrepancy of this half-life with the correct value, we credit Schuman \textit{et al.} with the first correct identification of $^{142}$Ba.

\subsection*{$^{143}$Ba}\vspace{-.85cm}
Wahl \textit{et al.} reported the first identification of $^{143}$Ba in 1962 in the article \textit{Nuclear-Charge Distribution in Low-Energy Fission} \cite{Wah62}. $^{143}$Ba was produced from $^{235}$U fission induced by thermal neutrons. The neutrons were produced from reactions of 10 MeV deuterons accelerated by the Washington University cyclotron on a beryllium target. The half-life of $^{143}$Ba was measured by timed separations of its daughters. ``The half-life value of Ba$^{143}$ obtained was (12.0$\pm$1.2) sec.'' This value is consistent with the presently accepted value of 11.5(2)~s. Hahn and Strassmann had speculated about the existence of a short-lived barium isotope ($<$ 0.5~m) \cite{Hah42} which was tentatively assigned to $^{143}$Ba by the Plutonium Project Records \cite{PPR51}.

\subsection*{$^{144}$Ba}\vspace{-.85cm}
Amarel \textit{et al.} observed $^{144}$Ba in 1967 as reported in their article \textit{Half Life Determination of Some Short-Lived Isotopes of Rb, Sr, Cs, Ba, La and Identification of $^{93,94,95,96}$Rb as Delayed Neutron Precursors by On-Line Mass-Spectrometry} \cite{Ama67}. $^{144}$Ba was produced by fission of $^{238}$U induced by 150 MeV protons from the Orsay synchrocyclotron. Isotopes were identified with a Nier-type mass spectrometer and half-lives were determined by $\beta$ counting. The measured half-life for $^{144}$Ba was listed in the main table with 11.4(25)~s which is consistent with the currently adopted value of 11.5(2)~s.

\subsection*{$^{145}$Ba}\vspace{-.85cm}
Grapengiesser \textit{et al.} reported the observation of $^{145}$Ba in \textit{Survey of short-lived fission products obtained using the isotope-separator-on-line facility at Studsvik} in 1974 \cite{Gra74}. $^{145}$Ba was produced and identified at the OSIRIS isotope-separator online facility at the Studsvik Neutron Research Laboratory in Nyk\"oping, Sweden. In the long table of experimental half-lives of many different isotopes the half-life of $^{145}$Ba is quoted as 4.2(5)~s. This value agrees with the currently adopted value of 4.31(16)~s. The previous observation of $\gamma$-ray transitions attributed to $^{145}$Ba were not sufficiently accurate and not based on firm mass assignment \cite{Hop71}.

\subsection*{$^{146}$Ba}\vspace{-.85cm}
The paper \textit{Ground-State Bands in Neutron-Rich Even Te, Xe, Ba, Ce, Nd, and Sm Isotopes Produced in the Fission of $^{252}$Cf} published in 1970 reported the first identification of $^{150}$Ce by Wilhelmy {\it et al.} at Berkeley \cite{Wil70}. They measured $\gamma$-spectra following spontaneous fission of $^{252}$Cf and observed the first two $\gamma$-ray transitions of the $^{146}$Ba ground state band. They did not mention the first observation of $^{146}$Ba: ``The data, which in some of the cases can be correlated with previously reported results...'' implies that some of the cases were new observations.

\subsection*{$^{147}$Ba}\vspace{-.85cm}
Wohn \textit{et al.} reported the discovery of $^{147}$Ba in 1978 in their article \textit{Identification of $^{147}$Cs and Half-Life Determinations for Cs and Ba Isotopes with A=144-147 and Rb and Sr Isotopes with A=96-98} \cite{Woh78}. $^{147}$Ba was produced and identified by neutron induced fission of $^{235}$U at the On-line Separator f\"{u}r Thermisch Ionisierbare Spaltprodukte (OSTIS) facility of the Institut Laue-Langevin in Grenoble, France. ``Half-life determinations of Rb and Cs fission products available at an on-line mass separator have been made for several neutron-rich Rb, Sr, Cs, and Ba isotopes using both $\beta$-multiscale and $\gamma$-multispectra measurements. The half-lives and rms uncertainties (in sec) are...$^{147}$Ba, 0.70(6).'' The observed value for the half-life is close to the accepted value of 0.893(1)~s. Wohn \textit{et al.} were aware of a previous work for $^{147}$Ba published in a conference proceeding \cite{Ami76}. This work was submitted by Engler \textit{et al.} to a refereed journal \cite{Eng79} seven months later than Wohn \textit{et al.}. Engler \textit{et al.} claimed the first observation of $^{147}$Ba although they quote the work by Wohn \textit{et al.}.

\subsection*{$^{148}$Ba}\vspace{-.85cm}
Engler \textit{et al.} observed $^{148}$Ba for the first time as reported in the 1979 article \textit{Half-Life Measurements of Rb, Sr, Y, Cs, Ba, La and Ce Isotopes with A=91-98 and A=142-149} \cite{Eng79}. A $^{235}$U target was exposed to thermal neutrons at the Soreq Nuclear Research Centre in Yavne, Israel. $^{148}$Ba was identified with the Soreq-On-Line-Isotope-Separator (SOLIS). ``The isotopes $^{147,148}$Ba and $^{149}$La were identified for the first time and their half-lives measured. The values obtained, in seconds, are 0.72$\pm$0.07 for $^{147}$Ba, 0.47$\pm$0.20 for $^{148}$Ba and 1.2$\pm$0.4 for $^{149}$La.'' The half-life for $^{148}$Ba is consistent with the accepted value of 0.612(17)~s.

\subsection*{$^{149}$Ba}\vspace{-.85cm}
In the 1993 article \textit{Delayed-neutron branching ratios of precursors in the fission product region} Rudstam \textit{et al.} reported the observation of $^{149}$Ba \cite{Rud93}. $^{149}$Ba was produced and identified at the OSIRIS isotope-separator online facility at the Studsvik Neutron Research Laboratory in Nyk\"oping, Sweden. In the large table of delayed-neutron branching ratios and half-lives the half-life of $^{149}$Ba is quoted as 0.324(18)~s. This value agrees with the currently adopted value of 344(7)~s. Warner and Reeder had reported a half-life measurement for $^{149}$Ba seven years earlier in a conference proceeding \cite{War86}.

\subsection*{$^{150-151}$Ba}\vspace{-.85cm}
Bernas {\it{et al.}} discovered $^{150}$Ba and $^{151}$Ba in 1994 at GSI, Germany, as reported in {\it{Projectile Fission at Relativistic Velocities: A Novel and Powerful Source of Neutron-Rich Isotopes Well Suited for In-Flight Isotopic Separation}} \cite{Ber94}. The isotopes were produced using projectile fission of $^{238}$U at 750 MeV/nucleon on a lead target. ``Forward emitted fragments from $^{80}$Zn up to $^{155}$Ce were analyzed with the Fragment Separator (FRS) and unambiguously identified by their energy-loss and time-of-flight.'' The experiment yielded 13 individual counts of $^{151}$Ba. As shown in Figure 3 of the article many more counts of $^{150}$Ba were recorded though not explicitly mentioned in the text since Mach \textit{et al.} had reported the discovery of $^{150}$Ba in a conference abstract \cite{Mac87}. However, since this observation was never published in a refereed journal we credit Bernas {\it et al.} with the discovery of $^{150}$Ba.

\section{Summary}
The discovery of the barium isotopes has been cataloged and the methods of their discovery discussed. Many of the barium isotopes have a long and interesting history. The discoveries of $^{140}$Ba and $^{141}$Ba were directly linked to the discovery of fission. The half-lives of four isotopes ($^{128}$Ba, $^{139}$Ba, $^{142}$Ba, and $^{143}$Ba) were first measured without accurate mass assignments and two measurements were initially wrong ($^{125}$Ba and $^{142}$Ba). $^{129}$Ba and $^{147}$Ba were identified essentially simultaneously by two groups independently. It is also interesting to note that $^{120}$Ba, $^{122}$Ba, $^{124}$Ba, and $^{146}$Ba were first observed in $\gamma$-ray spectroscopy studies. Finally, the discovery of $^{149}$Ba and $^{150}$Ba had been reported in conference proceedings seven years prior to a publication in refereed journals.

\ack

This work was supported by the National Science Foundation under grants No. PHY06-06007 (NSCL) and PHY07-54541 (REU). MH was supported by NSF grant PHY05-55445. JQG acknowledges the support of the Professorial Assistantship Program of the Honors College at Michigan State University.


\newpage

\section*{EXPLANATION OF TABLE}\label{sec.eot}
\addcontentsline{toc}{section}{EXPLANATION OF TABLE}

\renewcommand{\arraystretch}{1.0}

\begin{tabular*}{0.95\textwidth}{@{}@{\extracolsep{\fill}}lp{5.5in}@{}}
\textbf{TABLE I.}
	& \textbf{Discovery of Barium Isotopes }\\
\\

Isotope & Barium isotope \\
Author & First author of refereed publication \\
Journal & Journal of publication \\
Ref. & Reference \\
Method & Production method used in the discovery: \\
 & FE: fusion evaporation \\
 & LP: light-particle reactions (including neutrons) \\
 & MS: mass spectroscopy \\
 & PN: photonuclear reactions \\
 & SF: spontaneous fission \\
 & NF: neutron-induced fission \\
 & CPF: charged-particle induced fission \\
 & PF: projectile fragmentation or projectile fission \\
Laboratory & Laboratory where the experiment was performed\\
Country & Country of laboratory\\
Year & Year of discovery \\
\end{tabular*}
\label{tableI}

\newpage
\datatables

\setlength{\LTleft}{0pt}
\setlength{\LTright}{0pt}


\setlength{\tabcolsep}{0.5\tabcolsep}

\renewcommand{\arraystretch}{1.0}


\begin{longtable}[c]{%
@{}@{\extracolsep{\fill}}r@{\hspace{5\tabcolsep}} llllllll@{}}
\caption[Discovery of Barium Isotopes]%
{Discovery of Barium isotopes}\\[0pt]
\caption*{\small{See page \pageref{tableI} for Explanation of Tables}}\\
\hline
\\[100pt]
\multicolumn{8}{c}{\textit{This space intentionally left blank}}\\
\endfirsthead
Isotope & Author & Journal & Ref. & Method & Laboratory & Country & Year \\

$^{114}$Ba & A. Guglielmetti & Phys. Rev. C & Gug95 & FE & Darmstadt & Germany &1995 \\
$^{115}$Ba & Z. Janas & Nucl. Phys. A & Jan97 & FE & Darmstadt & Germany &1997 \\
$^{116}$Ba & Z. Janas & Nucl. Phys. A & Jan97 & FE & Darmstadt & Germany &1997 \\
$^{117}$Ba & D.D. Bogdanov & Nucl. Phys. A & Bog77 & FE & Dubna & Russia &1977 \\
$^{118}$Ba & Z. Janas & Nucl. Phys. A & Jan97 & FE & Darmstadt & Germany &1997 \\
$^{119}$Ba & D.D. Bogdanov & Sov. J. Nucl. Phys. & Bog74 & FE & Dubna & Russia &1974 \\
$^{120}$Ba & J. Conrad & Nucl. Phys. A & Con74 & FE & Heidelberg & Germany &1974 \\
$^{121}$Ba & D.D. Bogdanov & Sov. J. Nucl. Phys. & Bog75 & FE & Dubna & Russia &1975 \\
$^{122}$Ba & J. Conrad & Nucl. Phys. A & Con74 & FE & Heidelberg & Germany &1974 \\
$^{123}$Ba & I.L. Preiss & J. Inorg. Nucl. Chem. & Pre62 & FE & Yale & USA &1962 \\
$^{124}$Ba & J.E. Clarkson & Nucl. Phys. A & Cla67 & FE & Berkeley & USA &1967 \\
$^{125}$Ba & J.M. D'Auria & Phys. Rev. & DAu68 & FE & Yale & USA &1968 \\
$^{126}$Ba & M.I. Kalkstein & J. Inorg. Nucl. Chem. & Kal54 & FE & Berkeley & USA &1954 \\
$^{127}$Ba & M. Linder & Phys. Rev. & Lin52 & LP & Livermore & USA &1952 \\
$^{128}$Ba & R.W. Fink & Phys. Rev. & Fin50 & LP & Berkeley & USA &1950 \\
$^{129}$Ba & C.C. Thomas & Phys. Rev. & Tho50 & LP & Rochester & USA &1950 \\
$^{130}$Ba & A.J. Dempster & Phys. Rev. & Dem36 & MS & Chicago & USA &1936 \\
$^{131}$Ba & S. Katcoff & Phys. Rev. & Kat47 & NC & Argonne & USA &1947 \\
$^{132}$Ba & A.J. Dempster & Phys. Rev. & Dem36 & MS & Chicago & USA &1936 \\
$^{133}$Ba & J.M. Cork & Phys. Rev. & Cor41 & LP & Michigan & USA &1941 \\
$^{134}$Ba & J.P. Blewitt & Phys. Rev. & Ble36 & MS & Princeton & USA &1936 \\
$^{135}$Ba & F.W. Aston & Proc. Roy. Soc. & Ast32 & MS & Cambridge & UK &1932 \\
$^{136}$Ba & F.W. Aston & Proc. Roy. Soc. & Ast32 & MS & Cambridge & UK &1932 \\
$^{137}$Ba & F.W. Aston & Proc. Roy. Soc. & Ast32 & MS & Cambridge & UK &1932 \\
$^{138}$Ba & F.W. Aston & Phil. Mag. & Ast25 & MS & Cambridge & UK &1925 \\
$^{139}$Ba & M.L. Pool & Phys. Rev. & Poo37& NC & Michigan & USA &1937 \\
$^{140}$Ba & O. Hahn & Naturwiss. & Hah39 & NF & Berlin & Germany &1939 \\
$^{141}$Ba & S. Katcoff & Nat. Nucl. Ener. Ser. & Kat45 & NF & Chicago & USA &1945 \\
$^{142}$Ba & R.P. Schuman & Phys. Rev. & Sch59 & NF & Idaho Falls & USA &1959 \\
$^{143}$Ba & A.C. Wahl & Phys. Rev. & Wah62 & NF & St. Louis & USA &1962 \\
$^{144}$Ba & I. Amarel & Phys. Lett. B & Ama67 & CPF & Orsay & France &1967 \\
$^{145}$Ba & B. Grapengiesser & J. Inorg. Nucl. Chem. & HGra74 & NF & Studsvik & Sweden &1974 \\
$^{146}$Ba & J.B. Wilhelmy & Phys. Rev. Lett. & Wil70 & SF & Berkeley & USA &1970 \\
$^{147}$Ba & F.K. Wohn & Phys. Rev. C & Woh78 & NF & Grenoble & France &1978 \\
$^{148}$Ba & G. Engler & Phys. Rev. C & Eng79 & NF & Soreq & Israel &1979 \\
$^{149}$Ba & G. Rudstam & At. Data Nucl. Data Tables & Rud93 & NF & Studsvik & Sweden &1993 \\
$^{150}$Ba & M. Bernas & Phys. Lett. B & Ber94 & PF & Darmstadt & Germany &1994 \\
$^{151}$Ba & M. Bernas & Phys. Lett. B & Ber94 & PF & Darmstadt & Germany &1994 \\
\end{longtable}
\newpage


\normalsize

\begin{theDTbibliography}{1956He83}
\bibitem[Ama67]{Ama67t} I. Amarel, R. Bernas, R. Foucher, J. Jastrzebski, A. Johnson, J. Teillac, and H. Gauvin, Phys. Lett. {\bf 24B}, 402 (1967)
\bibitem[Ast25]{Ast25t} F.W. Aston, Phil. Mag. {\bf 49}, 1191 (1925)
\bibitem[Ast32]{Ast32t} F.W. Aston, Proc. Roy. Soc. A {\bf 134}, 571 (1932)
\bibitem[Ber94]{Ber94t} M. Bernas, S. Czajkowski, P. Armbruster, H. Geissel, Ph. Dessagne, C. Donzaud, H-R. Faust, E. Hanelt, A. Heinz, M. Hesse, C. Kozhuharov, Ch. Miehe, G. M\"unzenberg, M. Pf\"utzner, C. R\"ohl, K.-H. Schmidt, W. Schwab, C. St\'ephan, K. S\"ummerer, L. Tassan-Got, and B. Voss, Phys. Lett. B {\bf 331}, 19 (1994)
\bibitem[Ble36]{Ble36t} J.P. Blewett and M.B. Sampson, Phys. Rev. {\bf 49}, 778 (1936)
\bibitem[Bog74]{Bog74t} D.D. Bogdanov, V.A. Karnaukhov, and L.A. Petrov, Sov. J. Nucl. Phys. {\bf 19}, 940 (1974)
\bibitem[Bog75]{Bog75t} D.D. Bogdanov, A.V. Demyanov, V.A. Karnaukhov, and L.A. Petrov, Sov. J. Nucl. Phys. {\bf 21}, 233 (1975)
\bibitem[Bog77]{Bog77t} D.D. Bogdanov, A.V. Demyanov, V.A. Karnaukhov, L.A. Petrov, A. Plohocki, V.G. Subbotin, and J. Voboril, Nucl. Phys. A {\bf 275}, 229 (1977)
\bibitem[Cla67]{Cla67t} J.E. Clarkson, R.M. Diamond, F.S. Stephens, and I. Perlman, Nucl. Phys. A {\bf 93}, 272 (1967)
\bibitem[Con74]{Con74t} J. Conrad, R. Repnow, E. Grosse, H. Homeyer, E. Jaeschke,and J.P. Wurm, Nucl. Phys. A {\bf 234}, 157 (1974)
\bibitem[Cor41]{Cor41t} J.M. Cork and G.P. Smith, Phys. Rev. {\bf 60}, 480 (1941)
\bibitem[DAu68]{DAu68t} J.M. D'Auria, H. Bakhru, and I.L. Preiss, Phys. Rev. {\bf 172}, 1176 (1968)
\bibitem[Dem36]{Dem36t} A.J. Dempster, Phys. Rev. {\bf 49}, 947 (1936)
\bibitem[Eng79]{Eng79t} G. Engler, Y. Nir-El, M. Shmid, and S. Amiel, Phys. Rev. C {\bf 19}, 1948 (1979)
\bibitem[Fin50]{Fin50t} R.W. Fink and D.H. Templeton, Phys. Rev. {\bf 72}, 2818 (1950)
\bibitem[Gra74]{Gra74t} B. Grapengiesser, E. Lund, and G. Rudstam, J. Inorg. Nucl. Chem. {\bf 36}, 2409 (1974)
\bibitem[Gug95]{Gug95t} A. Guglielmetti, R. Bonetti, G. Poli, P. B. Price, A. J. Westphal, Z. Janas, H. Keller, R. Kirchner, O. Klepper, A. Piechaczek, E. Roeckl, K. Schmidt, A. Plochocki, J. Szerypo, and B. Blank, Phys. Rev. C {\bf 52}, 740 (1995)
\bibitem[Hah89]{Hah39t} O. Hahn and F. Strassman, Naturwiss. {\bf 27}, 89 (1939)
\bibitem[Jan97]{Jan97t} Z. Janas, A. Plochocki, J. Szerypo, R. Collatz, Z. Hu, H. Keller, R. Kirchner, O. Klepper, E. Roeckl, K. Schmidt, R. Bonetti, A. Guglielmetti, G. Poli, A. Piechaczek, Nucl. Phys. A {\bf 627}, 119 (1997)
\bibitem[Kal54]{Kal54t} M.I. Kalkstein and J.M. Hollander, Phys. Rev. {\bf 96}, 730 (1954)
\bibitem[Kat45]{Kat45t} S. Katcoff {\it Radiochemical Studies: The Fission Products}, Paper 172, p. 1147, National Nuclear Energy Series IV, 9, (McGraw-Hill, New York 1951)
\bibitem[Kat47]{Kat47t} S. Katcoff, Phys. Rev. {\bf 72}, 1160 (1947)
\bibitem[Lin52]{Lin52t} M. Linder and R.N. Osborne, Phys. Rev. {\bf 88}, 1422 (1952)
\bibitem[Poo37]{Poo37t} M.L. Pool, J.M. Cork, and R.L. Thornton, Phys. Rev. {\bf 52}, 239 (1937)
\bibitem[Pre62]{Pre62t} I.L. Preiss, and P.M. Strudler, J. Inorg. Nucl. Chem. {\bf 24}, 589 (1962)
\bibitem[Rud93]{Rud93t} G. Rudstam, K. Aleklett, and L. Sihver, At. Data. Nucl. Data. Tables {\bf 53}, 1 (1993)
\bibitem[Sch59]{Sch59t} R.P. Schuman, E.H. Turk, and R.L. Heath, Phys. Rev. {\bf 115}, 185 (1959)
\bibitem[Tho50]{Tho50t} C.C. Thomas and E.O. Wiig, Phys. Rev. {\bf 72}, 2818 (1950)
\bibitem[Wah62]{Wah62t} A.C. Wahl, R. L. Ferguson, D.R. Nethaway, D.E. Troutner, and K. Wolfsberg, Phys. Rev. C {\bf 126}, 1112 (1962)
\bibitem[Wil70]{Wil70t} J.B. Wilhelmy, S.G. Thompson, R.C. Jared, and E. Cheifetz, Phys. Rev. Lett. {\bf 25}, 1122 (1970)
\bibitem[Woh78]{Woh78t} F.K. Wohn, K.D. Wunsch, H. Wollnik, R. Decker, G. Jung, E. Koglin, and G. Siegert, Phys. Rev. C {\bf 17}, 2185 (1978)

\end{theDTbibliography}

\end{document}